\documentclass[a4,twocolumn]{aa}
\newcommand{\mr}{\mathrm}
\usepackage{graphicx,txfonts,natbib,verbatim}
\bibpunct{(}{)}{;}{a}{}{,}

\begin{document}

\title{The spectral difference between solar flare HXR coronal and footpoint
sources due to wave-particle interactions.}

\author{I. G. Hannah \and E. P. Kontar}

\offprints{Hannah \email{iain@astro.gla.ac.uk}}

\institute{School of Physics \& Astronomy, University of Glasgow, Glasgow, G12
8QQ, UK}

\date{Received ; Accepted }

\abstract{}{Investigate the spatial and spectral evolution of hard X-ray (HXR)
emission from flare accelerated electron beams subject to collisional transport
and wave-particle interactions in the solar atmosphere.}{We numerically follow
the propagation of a power-law of accelerated electrons in  {1D} space
and time with the response of the background plasma in the form of Langmuir
waves using the quasilinear approximation.}{We find that the addition of
wave-particle interactions to collisional transport for a transient initially
injected
electron beam flattens the spectrum of the footpoint source.  The coronal source
is unchanged and so the difference in the spectral indices between the coronal
and footpoint sources is $\Delta \gamma > 2$, which is larger than expected
from purely collisional transport. A steady-state beam shows little difference
between the two cases, as has been previously found, as a transiently injected
electron beam is required to produce significant wave growth, especially at
higher velocities. With this transiently injected beam the wave-particle
interactions dominate in the corona whereas the collisional losses dominate in
the chromosphere. The shape of the spectrum is different with increasing
electron beam density in the wave-particle interaction case whereas with purely
collisional transport only the normalisation is changed. We also find that the
starting height of the source electron beam above the photosphere affects the
spectral index of the footpoint when Langmuir wave growth is included. This
may account for the differing spectral indices found between double footpoints
if asymmetrical injection has occurred in the flaring loop.}{}


\titlerunning{HXR spectral difference due to wave-particle interactions.}

\authorrunning{Hannah et al.}

\maketitle

\section{Introduction}
Solar flares are prolific accelerators of electrons, sending beams outwards into
interplanetary space and downwards deeper into the solar atmosphere. The
latter is inferred from radio and hard X-ray  (normally above $> 20$ keV, HXR)
observations which suggest coronally accelerated electrons propagating
downwards along magnetic field lines towards the denser chromosphere,
eventually stopping via Coulomb collisions heating the local plasma
\citep[e.g.][]{brown1971,2003ApJ...595L.115B}. This heated chromospheric
material expands upwards filling the magnetic loops, producing bright emission
in soft X-rays SXR and EUV. In this standard model a power-law of accelerated
electrons in the corona $F_\mathrm{0}(E) \propto E^{-\delta_\mathrm{b}}$ only
encounters Coulomb collisions with the background plasma whilst it propagates
to the chromosphere. Assuming a time and spatially independent electron
injection, the footpoint source will emit HXR with spectrum $I(\epsilon)\propto
\epsilon^{-\gamma_\mathrm{FP}}$ where
$\gamma_\mathrm{FP}=\delta_\mathrm{b}-1$, namely the thick target model
\citep{brown1971,1972SvA....16..273S}. The same beam of accelerated electrons
should continuously emit via bremsstrahlung with the background plasma as it
propagates from the corona. This thin-target emission (assuming no energy loss
by the electrons)
\citep{1968ApJ...151..711A,1968ApJ...154.1027H,1969SoPh....6..133T,
1976SoPh...50..153L} is again found to produce a power-law spectrum but this
time with an index of $\gamma_\mathrm{CS}=\delta_\mathrm{b}+1$. So if the
HXR emission of the electron beam was observable from the corona to the
chromosphere then the acceleration and transport processes involved could be
well constrained.

However the footpoint emission dominates observations due to the properties of
the bremsstrahlung emission process, namely an electron distribution $F(E,r)$
produces a photon flux emission $I(\epsilon)$, in units of photons cm$^{-2}$
keV$^{-1}$ s$^{-1}$, of

\begin{equation}\label{eq:brem} I(\epsilon)\propto \int_\epsilon^\infty
\int_V
n(r)F(E,r)Q(\epsilon,E) dE d^3r
\end{equation}

\noindent where $n(r)$ is the background plasma density and $Q(\epsilon,E)$ is
the bremsstrahlung cross-section
\citep{1959RvMP...31..920K,1997A&A...326..417H}. Given that the
chromospheric density is several orders of magnitude greater than the corona
this emission dominates the observed spatially integrated HXR spectrum. HXR
imaging spectroscopy instruments, such as Yohkoh/HXT
\citep{1992PASJ...44L..45K}, RHESSI \citep{2002SoPh..210....3L}, have a limited
dynamic range and so the faint coronal emission is rarely observed with the
bright footpoints. However observing both would be very effective for
understanding the physics of transport and acceleration of electrons in flares.

The standard flare model predicts the difference between the observed spectral
indices of the coronal and footpoint sources should be

\begin{equation}\label{eq:diff}
\Delta \gamma=\gamma_\mathrm{CS}-\gamma_\mathrm{FP}=2.
\end{equation}

\noindent While there are solar flare events with the predicted spectral index
difference the majority of observations do not demonstrate this behaviour. The
first HXR observations of coronal and footpoint sources were made by
\citet{1994Natur.371..495M} with Yohkoh/HXT and the ``coronal source above
two footpoints'' geometry of this event has dominated subsequent models. The
spectral indices of the coronal and footpoint sources were found by taking the
ratio of images in $14-23$~keV to $23-33$~keV and $23-33$~keV to
$33-53$~keV energy bands finding $\gamma_\mathrm{FP}=2.0$, 4.0 and
$\gamma_\mathrm{CS}=2.6$, 4.1 respectively \citep{1995PASJ...47..677M}. Thus
finding virtually no difference between the coronal and footpoints sources and
certainly not $\Delta \gamma=2$. This event was further analysed and found to
be inconsistent with thin-target coronal emission and thick-target footpoints
\citep{1997ApJ...489..442A}.  Using an improved analysis technique
\citet{2000AdSpR..26..493M} found that the difference in spectral indices was
larger than previously thought but still $\Delta \gamma<2$. They also suggest
such coronal emission may not be non-thermal in origin but possibly due to a
very hot (100MK) thermal source.  {RHESSI observations of such
``above-the-loop-top sources'' demonstrate power-law and not thermal spectra
\citep[e.g.][]{2010ApJ...714.1108K}.} A survey of Yohkoh flares with
simultaneous coronal and footpoint emission found that the coronal emission
was typically softer yet the average difference was only $\Delta
\gamma=1.3\pm 1.5$ \citep{2002ApJ...569..459P}.

With RHESSI, spatially resolved spectroscopy with high energy resolution
became routinely available and triggered a number of interesting findings
\citep{2003ApJ...595L.107E,2007A&A...466..713B,2008SoPh..250...53S,
2008A&A...487..337B,2009ApJ...705.1584S}. \citet{2006A&A...456..751B}
analysed five flares obtaining HXR spectra of simultaneous footpoint and
coronal sources finding a range of $0<\Delta \gamma <6$. Further investigation
of three of these flares with clearly separated footpoint and coronal sources
found that $\Delta \gamma > 3$. Their explanation for the discrepancy is that
there was an anomalously large density concentrated at the coronal source so
that this emission is a combination of thin and thick-target, which would result
in a flatter footpoint spectral index at low energies and $\Delta \gamma
>
2$ \citep{2007A&A...466..713B}. \citet{1995SoPh..158..283W} had previously
suggested this solution to the coronal sources observed by Yohkoh/HXT
\citep{1994Natur.371..495M,1994ApJ...421..843F}. However this model predicts
a break in the coronal spectrum between the thin and thick target range which
has not been observed with RHESSI despite its high energy resolution. A further
suggestion was the inclusion of non-collisional losses implemented via a
decelerating electric potential \citep{2008A&A...487..337B}. This electric field
was assumed to be due to the return current generated by the electron beam
and was implemented in a time-independent manner. It was found that this did
flatten the footpoint spectrum, providing the necessary larger difference
between spectral indices, however the simple empirical implementation had
problems with the relative brightness of the sources, in one event producing
considerably fainter footpoint sources than observed.

An additional constraint can be gained by considering the difference in spectral
indices between the footpoint sources themselves (normally two footpoints are
observed). In the standard flare model two similar footpoints are expected to
have similar spectral indices. However observations with RHESSI have shown
this not to be always the case. \citet{2003ApJ...595L.107E} analysed a large
X-class flare and found that during the event the two time correlated footpoint
sources differed in spectral index with $\Delta\gamma_\mathrm{FP}\approx 0.3
- 0.4$. \citet{2006A&A...456..751B} found that the difference between the
footpoints' spectral index ranged from $0<\Delta\gamma_\mathrm{FP}<1$.
\citet{2008SoPh..250...53S} analysed 53 flares with a double footpoint structure
and found that the spectral index between them was
$0<\Delta\gamma_\mathrm{FP}<0.6$. An explanation for these results is that
collisional transport encountered different column densities in each leg of the
flaring loop \citep{2003ApJ...595L.107E}, however \citet{2008SoPh..250...53S}
demonstrated that this would result in the softer footpoint being brighter, the
opposite to observations. They instead prefer a magnetic mirroring model
\citep{1976MNRAS.176...15M} in which the magnetic field converges more
quickly at one of the footpoints. This convergence of the magnetic field lines
has been found to be consistent with area measurements of HXR footpoints
\citep{2006SPD....37.1308S} and has been observed in one M-class limb event
with RHESSI \citep{2008A&A...489L..57K}.

\begin{figure}\centering
\includegraphics[width=\columnwidth]{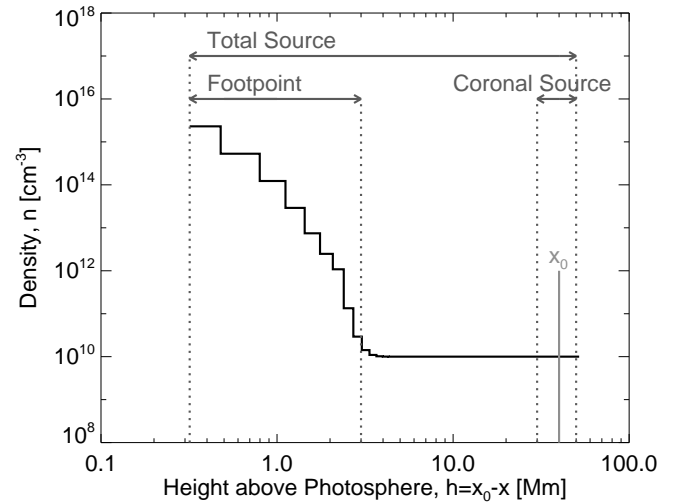}
\caption{\label{fig:n0} The
background plasma density $n(x)$ as a function of
height above the photosphere. The vertical lines indicate the
different spatial region the emission is summed over to produce the total,
footpoint and coronal sources. The short grey vertical line $x_\mr{0}$ indicates
a starting centroid height of the electron distribution at 40Mm.} \end{figure}

In both cases the differences in spectral indices is not what is expected from
purely collisional transport and something more is needed. In this paper we
consider the generation of Langmuir waves via wave-particle interactions as the
response of the background plasma to the propagating electron beam, in
addition to Coulomb collisions. The observation of downward moving decimetric
radio emission in some flares -- Reverse-Slope Type III bursts --
\citep[e.g.][]{1975Natur.258..693T,1995ApJ...455..347A,1997A&A...320..612K,
1997ApJ...480..825A} is thought to be a signature of Langmuir waves being
generated by the downward moving electron beam. The presence of Langmuir
waves is inferred as they are the starting point for non-linear processes that
result in this non-thermal radio emission. Although the majority of flares do
not
have such emission \citep{2005SoPh..226..121B} it does not mean Langmuir
waves are not present but that the radio emission itself has been absorbed .
Previous studies of this fast non-collisional process in flares have mostly
focused on time and spatially independent solutions
\citep{1984ApJ...279..882E,hamilton1987,McClements1987}. These steady-state
studies found that while the distribution of electrons changes due to the
generation of Langmuir waves, the spatially integrated spectra shows
insignificant changes over purely collisional transport.

\begin{figure*}\centering
\includegraphics[width=1.8\columnwidth]{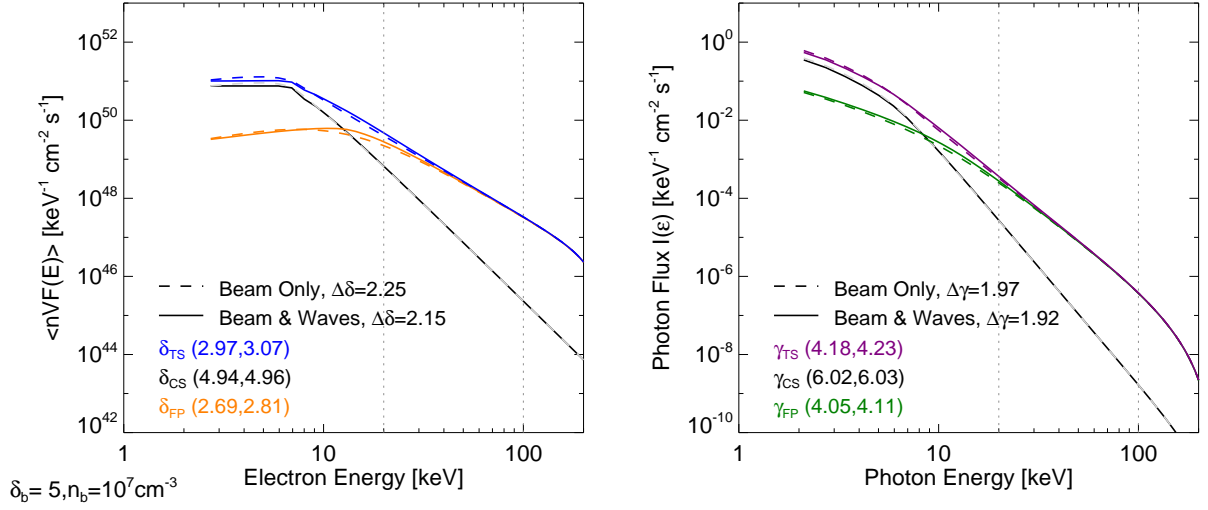}
\caption{\label{fig:conspec} Mean electron flux spectrum (left) and X-ray
spectrum (right) of continually injected electron beam simulation after reaching
equilibrium (steady-state), for the footpoint (orange, green), coronal (black)
and
total source (blue, purple). Shown are the simulations considering only
collisional transport of an electron beam (\emph{Beam Only}, dashed lines) and
the addition of wave-particle interactions (\emph{Beam \& Waves}, solid lines).
The vertical dotted lines indicated the energy range a power law was fitted
over.
The fitted spectral indices are shown in brackets (\emph{Beam Only},
\emph{Beam \& Waves}).}
\end{figure*}

However, a highly intermittent and bursty injection of electrons is more
realistic
and often observed
\citep{Kiplinger_etal1984,1994SoPh..153..403F,Aschwanden_etal1998} and
spatial filamentation (non-monolithic multi-thread flaring loops)  {can}
explain simultaneously the heights and the sizes of observed HXR footpoints
\citep{2010ApJ...717..250K}. So to be able to simulate the propagation of
 {such a possibly} fragmented (temporally and spatially) electron
acceleration it is crucial to consider the time dependent injection and
transport
of electrons. As \citet{2009ApJ...707L..45H} have previously shown in this
situation the generation of Langmuir waves can have a strong affect on the
electron distribution and the observed HXR spectrum.

In \S \ref{sec:setup} we describe the system of equations we are numerically
solving to follow the propagation of the accelerated electrons and the response
of the background plasma in the form of Langmuir waves. In \S \ref{sec:con} we
demonstrate that the inclusion of wave-particle interactions of a continually
injected electron beam that has reached a steady-state has little effect on the
electron distribution, as has been previously noted by other authors
\citep{hamilton1987}. In \S \ref{sec:inst} and \S \ref{sec:diff_cf} we show that
for a transiently injected beam the wave-particle interactions do have an effect
and show how this changes the difference in spectral indices between the
coronal and footpoint sources. In \S \ref{sec:height} we show that the starting
height of the beam effects the growth of Langmuir waves and the hardness of
the resulting footpoint spectrum.

\section{Electron Beam Simulation}\label{sec:setup}

We describe the 1D electron velocity ($v\approx v_{||} \gg v_\bot$) transport of
an electron beam $f(v,x,t)$ [electrons cm$^{-4}$ s]  from the corona to
chromosphere, self-consistently driving Langmuir waves (of spectral energy
density $W(v,x,t)$ [erg cm$^{-2}$]) using the equations of quasi-linear
relaxation \citep{VedenovVelikhov1963,DrummondPines1964,
1969JETP...30..131R,hamilton1987,2001SoPh..202..131K,2009ApJ...707L..45H}.
This weakly turbulent description does not require resolving smallest scales
comparable to the plasma period/Debye length \citep[c.f. PIC simulations,
e.g.][]{2009A&A...506.1437K}, so the electron distribution can be followed at a
variety of realistic heights in the solar atmosphere. We consider several
processes in addition to the quasilinear wave-particle interaction -- Coulomb
collisions, Landau damping and spontaneous emission -- resulting in our
quasilinear equations becoming

\begin{equation}\label{eq:f} \frac{\partial f}{\partial t} +v\frac{\partial
f}{\partial x}= \frac{4\pi^2 e^2}{m^2} \frac{\partial}{\partial v}\left(
\frac{W}{v}\frac{\partial f}{\partial
v}\right)+\gamma_\mathrm{C_F}\frac{\partial}{\partial
v}\left(\frac{f}{v^2}\right) \end{equation}

\begin{equation}\label{eq:w} \frac{\partial W}{\partial t}+
\frac{3v_\mathrm{T}^2}{v}\frac{\partial W}{\partial x} = \left( \frac{\pi
\omega_\mathrm{p}}{n}v^2 \frac{\partial f}{\partial
v}-\gamma_\mathrm{C_W}-2\gamma_\mathrm{L} \right)W+Sf, \end{equation}

\noindent where $n(x)$ the background plasma density and
$\omega_\mathrm{p}(x)^2=4\pi n(x) e^2/m$ is the local plasma frequency. The
resonant interaction between the electrons and Langmuir waves,
$\omega_\mr{p_e}=kv$, is handled by the quasilinear term, the first term on the
righthand side of Equations \ref{eq:f} and \ref{eq:w}. The Coulomb collisions
terms for the electrons is $\gamma_\mathrm{C_F}=4\pi e^4
n\ln{\Lambda}/m^2$ \citep{1981phki.book.....L,1978ApJ...224..241E} and waves
$\gamma_\mathrm{C_W}= 1/3\sqrt{2/\pi} \gamma_\mathrm{C_F}\simeq \pi
e^4n \ln{\Lambda}/(m^2 v_\mathrm{T}^3)$
\citep{1981phki.book.....L,1980panp.book.....M}. Where
$\ln{\Lambda}=\ln{(8\times 10^6n^{-1/2}T)}$ is the Coulomb logarithm, $T$ is
the temperature of the background plasma (taken as 1MK) and
$v_\mathrm{T}=\sqrt{k_\mathrm{B}T/m}$ is the velocity of a thermal electron,
$k_\mathrm{B}$ is the Boltzmann constant. Equation \ref{eq:f} without the
quasilinear term describes the propagation of the electrons subject only to
Coulomb collisions with the background thermal plasma. In Equation \ref{eq:w}
there is also the Landau damping rate
$\gamma_\mathrm{L}=\sqrt{\pi/8}\omega_\mathrm{p}\left(v/v_\mathrm{T}
\right)^3\exp{ \left(-v^2/2v_\mathrm{T}^2\right ) }$
\citep{1981phki.book.....L}.
The spontaneous emission rate is given by $S=\omega_\mathrm{p}^3m
v\ln{\left(v/v_\mathrm{T}\right)}/(4\pi n)$ which agrees with that of
\citet{1980panp.book.....M,Tsytovich1995,hamilton1987}.

The evolution of energetic electrons is considered in the energy domain above
3~keV, which is the low energy limit of RHESSI. The initial electron
distribution is
taken to be a broken power-law in velocity which is flat below the break
$v_\mathrm{C}=3.38\times10^{9}$cms$^{-1}$
($E_\mathrm{C}(v_\mathrm{C})=7$keV) and above it with index of
$2\delta_\mathrm{b}$ (hence spectral index of $\delta_\mathrm{b}$ in energy
space) i.e.

\begin{equation}\label{eq:init} f(v,x,t=0)\propto  n_\mathrm{b}
\exp{\left(-\frac{x^2}{d^2}\right)} \left\{ \begin{array}{l l} 1 & \quad
\mbox{if
$v<v_\mr{C}$}\\ \left( v/v_\mr{C}\right)^{-2\delta_\mathrm{b}} & \quad
\mbox{if $v \ge
v_\mr{C}$}\\ \end{array}\right.\end{equation}

\begin{figure*}\centering
\includegraphics[width=1.8\columnwidth]{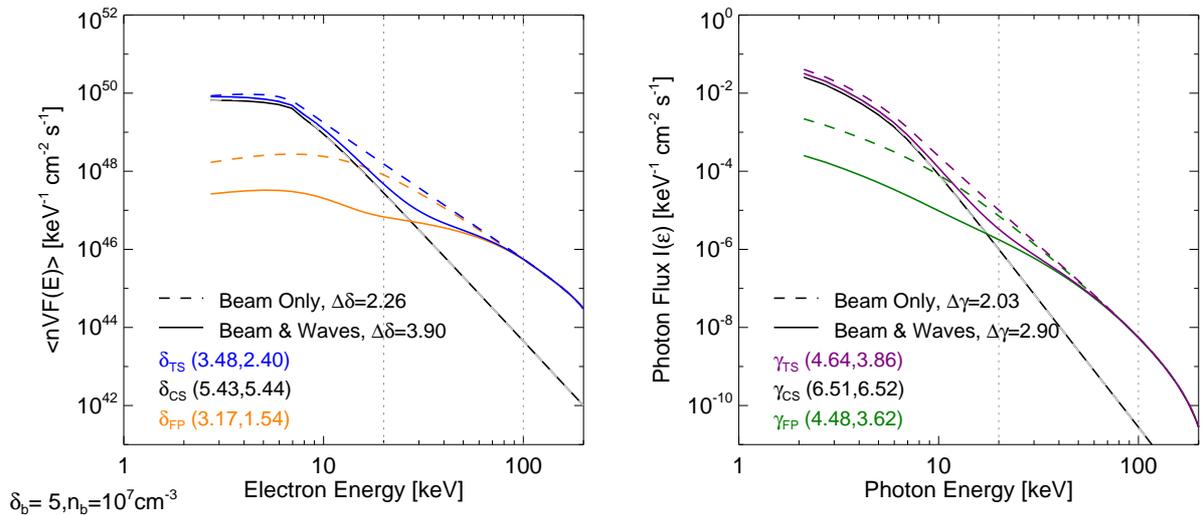}
\caption{\label{fig:inspec}Mean electron flux spectrum (left) and X-ray spectrum
(right) of a transient initially injected electron beam simulation over
1~second,
for the footpoint (orange, green), coronal (black) and total source (blue,
purple).
Shown are the simulations considering only collisional transport of an electron
beam (\emph{Beam Only}, dashed lines) and the addition of wave-particle
interactions (\emph{Beam \& Waves}, solid lines). The vertical dotted lines
indicated the energy range a power law was fitted over. The fitted spectral
indices are shown in brackets (\emph{Beam Only}, \emph{Beam \& Waves}.)}
\end{figure*}

\noindent where $n_\mathrm{b}$ is the beam density, $d$ is the characteristic
spatial Gaussian size. The simulation grid in x-space starts at 52Mm down to 0.3
Mm above the photosphere. Over this range the background density is at a
constant $10^{10}$cm$^{-3}$ in the corona which then steeply rises at the
transition region, exponentially increasing through the chromosphere, see Figure
\ref{fig:n0}. The electron distribution is initially centred at a height of
$x_\mathrm{0}$=40Mm with $d=2$~Mm. In the simulations beam densities of
$n_\mathrm{b}=10^7$cm$^{-3}$ and $n_\mathrm{b}=10^8$cm$^{-3}$ are used,
meaning that the initial number of electrons in this simulation is $N\approx 23
n_\mathrm{b}d^3=10^{33}-10^{34}$ electrons. This is an approximation as we
only have one spatial dimension to estimate the volume from, taking this as the
$FWHM=2\sqrt{2\ln{2}}d$. Compared to observations this is a modest number
of electrons to be accelerated, typical for a B or C-class flare, however
multiple
bursty injections of such beams could achieve the numbers derived for the
largest events. The initial wave spectral energy density is calculated from the
thermal background level.

In velocity space the simulation grid goes from $v_\mr{min}=7v_\mr{T}$ up to a
maximum of $1.05v_\mr{0}$, $v_\mr{0}$ the maximum beam velocity. For the
simulations shown $v_\mr{0}=75v_\mr{T}$ and so all over velocities are $v<c$.
Previous work on electron beam transport is predominantly non-relativistic
despite including velocities greater than $c$ \citep{hamilton1987} or energies
up to infinity \citep{2009ApJ...696..941S}. With our upper velocity limit we
remain in the non-relativistic regime but we will return to this later in
\S\ref{sec:maxv}.

We use a finite difference method to numerically solve Equations \ref{eq:f} and
\ref{eq:w} \citep{2001CoPhC.138..222K}. This code is written in a modular way
so that we can turn off transport processes and easily consider either electron
transport subject only to Coulomb collisions or with the additional
wave-particle
interactions. For collisional transport only we solve on Equation \ref{eq:f} but
without the first term on the righthand side. As there are no waves this is the
\emph{Beam Only} simulations. Considering both the collisional transport and
wave-particle interactions we solve both Equations \ref{eq:f} and \ref{eq:w} and
this is termed the \emph{Beam and Waves} case. This modular nature also
means that we can use an adaptive time-step in our code depending on the
processes being used. The quasilinear relaxation is predominantly the fastest
process,  occurring on a time-scale of $\tau_\mathrm{Q}= n/
(n_\mathrm{b}\omega_\mathrm{p})\approx 2\times10^{-5}
\sqrt{n}/n_\mathrm{b}$ seconds, where as the Coulomb collisions is
$\tau_\mathrm{C}=1/\gamma_\mathrm{C}\approx 1.5\times 10^7/n$ seconds.
For beam density $n_\mathrm{b}=10^8$cm$^{-3}$ the background density
would need to be $n>10^{14}$cm$^{-3}$ for the collisional time-scale to be
bigger than the quasilinear relaxation, a region electrons with sub-relativistic
energies rarely reach.

With these simulations we present two schemes of injecting the initial electron
distribution. First, shown in \S\ref{sec:con}, we have the continuous injection
of
the initial electron beam, this acting as a boundary condition until we reach a
steady-state after about 1.5 seconds simulation time. The other is a transient
initial injection of the electron beam which is not replenished, shown in
\S\ref{sec:inst}, which we numerically follow for 1 seconds in simulation time,
so that it will lose energy, completely leaving the simulation grid.

The electron beam will be instantaneously emitting X-rays $I(\epsilon)$ via
bremsstrahlung as given by Equation \ref{eq:brem}. For our simulated electron
$f(v,x,t)=F(E,x,t)/m_\mr{e}$ distribution this will be

\begin{equation}\label{eq:ph} I(\epsilon,x,t)=\frac{A}{4\pi R^2}
\left[n(x)\frac{f(v,x,t)}{m_\mr{e}}Q(\epsilon,E) \right]dE dx \end{equation}

\noindent where $A$ is the area of the emitting plasma. We take this to be
$FWHM^2=8(\ln{2})d^2$, and for the chosen $d$ the area matches the
characteristic sizes of flaring magnetic loops observed with RHESSI,
\citep[e.g.][]{2003ApJ...595L.107E,2010ApJ...717..250K}.

The time averaged spectrum from a particular spatial region of our simulation
(between $x_1$ and $x_2$) is

\begin{equation}\label{eq:ph_sum}
I(\epsilon)=\frac{A}{4\pi R^2} \sum_{E}
\sum_{x=x_1}^{x_2}
\sum_{t=0}^{t_f}\left[n(x)\frac{f(v,x,t)}{m_\mr{e}} Q(\epsilon,E) \right]dE dx
dt \end{equation}

\noindent This allows us to calculate the X-ray emission either from the total
source, coronal source (above 30Mm) and the footpoint (below 3Mm) as shown
in Figure~\ref{fig:n0}.

Another quantity of interest is the mean electron flux spectrum $\langle nVF(E)
\rangle$ as it is deducible from HXR spectrum \citep[e.g.][]{Brown_etal2006}.
From our simulations this is calculated as

\begin{equation}\label{eq:nvf}
\langle nVF(E) \rangle=\frac{A}{4\pi R^2} \sum_{x=x_1}^{x_2}
\sum_{t=0}^{t_f}\left[n(x)\frac{f(v,x,t)}{m_\mr{e}} \right]dx dt
\end{equation}

\noindent This is model independent value which can be inferred directly from
HXR observations.

\begin{figure*}\centering
\includegraphics[width=1.8\columnwidth]{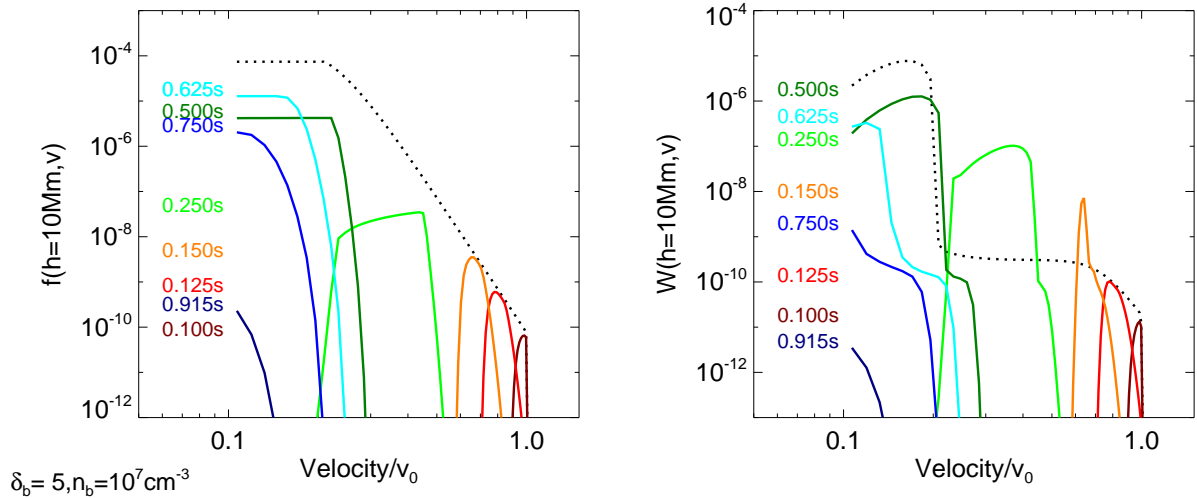}
\caption{\label{fig:fxv1010}Electron $f(v)$ and spectral wave energy $W(v)$
distributions for height of 10Mm above the photosphere. The dotted line
indicates the distributions for the continuous injection case once a
steady-state
equilibrium has been reached (shown in Figure \ref{fig:conspec}) and the solid
lines show the time evolution (different colours for different times as
indicated)
for the transient initially injected electron beam (shown in Figure
\ref{fig:inspec}).} \end{figure*}

\section{Simulation Results}
\subsection{Continuous Injection}\label{sec:con}

The first situation to consider is that of the steady-state continuous injection
of
the electron beam. This allows the most direct comparison to previous work on
purely collisional \citep{brown1971} and collisional plus wave-particle
interactions \citep{hamilton1987} transport, though in both cases they consider
a spatially independent solution. In our simulation we continuously inject a
beam (using $\delta_\mathrm{b}=5$ and $n_\mathrm{b}=10^7$cm$^{-3}$ in
Equation \ref{eq:init}) at the coronal source until a steady state is reached,
after
about 1.5 seconds. So observed HXR variations on timescales less than this are
likely due to non-stationary injection. The resulting steady-state mean electron
flux $\langle nVF(E) \rangle$ and X-ray spectrum $I(\epsilon)$ are shown in
Figure \ref{fig:conspec}.

For the coronal source emission the spectra are virtually identical, which
should
be expected since the transport processes should have little opportunity to act
on the distribution at this source location. Fitting a power-law over 20 - 100
keV
we obtain the expected $\delta_\mathrm{CS}\approx 5=\delta_\mathrm{b}$ in
the mean electron spectrum and $\gamma_\mathrm{CS}\approx
6=\delta_\mathrm{b}+1$ in the X-ray spectrum for the coronal source. We
choose this energy range to fit over as it is approximately over this range that
the power-law fit is performed in HXR imaging spectroscopy. At lower energies
the emission is predominantly thermal and at higher, background noise. For the
footpoint there is a slight difference between the electron beam (collisional
transport only) and beam and waves (collisions and wave-particle interactions)
simulation below about 30 keV but this is very minor. This confirms the previous
results that suggested that the addition of wave-particle interactions would
have little noticeable affect on the final spectra \citep{hamilton1987}. We also
achieve a spectral index of
$\gamma_\mathrm{FP}\approx4=\delta_\mathrm{b}-1$, resulting in $\Delta
\gamma \approx 2$, both as predicted. The total source spectra are slightly
softer than the thick-target prediction but that is because they include a faint
coronal source as well.

\subsection{Transient Injection}\label{sec:inst}

The second set of simulations have the same set up as the previous case
\S\ref{sec:con} except that we only have the injected electron beam at the start
and we run the simulation until the electrons have completely propagated
through the system, losing energy and thermalizing to the background
distribution and leave the numerical grid. This case is therefore referred to as
a
transient initial injection. The mean electron flux and X-ray spectra for these
simulations, integrated over the 1 second of simulation time it takes the
electrons to completely leave, are shown in Figure \ref{fig:inspec}. For the
collisional transport case (beam only) there are only minor differences with the
previous steady-state simulations. The magnitude of the simulation spectra are
lower in this case but that is because previously the electrons were being
continuously pumped into the system where as here it was only done initially.
The other difference is that the simulation spectra are all steeper (softer)
than
predicted by both the thin- and thick-target models. These however are
steady-state models and we have simulated the temporal evolution of the
electron beam so would expect some differences. Although we still obtain a
difference in spectral index of $\Delta \gamma \approx 2$ for the collisional
transport case.

With the addition of wave-particle interactions resulting in  Langmuir wave
growth we get a dramatically different spectrum for the footpoints (and hence
total source as well), considerably flatter than that found in the purely
collisional case. As the coronal source is again similar in both cases we obtain
a
difference in the spectral index bigger than the collisional case, the addition
of
wave-particle interactions producing $\Delta \gamma \approx 3$. In this case
the X-ray spectrum does start to deviate from a power-law in the range over
which we are fitting, in particular starting to flatten at lower energies. This
appearance of a break in the footpoint spectrum is consistent with those
observed in HXR spectroscopy but as we are comparing our simulation to HXR
imaging spectroscopy in which the energy resolution and noise make it
unsuitable to fit anything other than a single power-law
\citep{2006A&A...456..751B}, we will continue to use a single power-law.
Additionally the appearance of a break is exacerbated by the sharp drop off in
the spectra at the highest energies, beyond our fitting range. This is due to
the
upper velocity cut-off in the initial electron beam and the simulation grid and
we
will return to discuss this in detail in \S\ref{sec:maxv}.

\begin{figure*}\centering
\includegraphics[width=0.9\columnwidth]{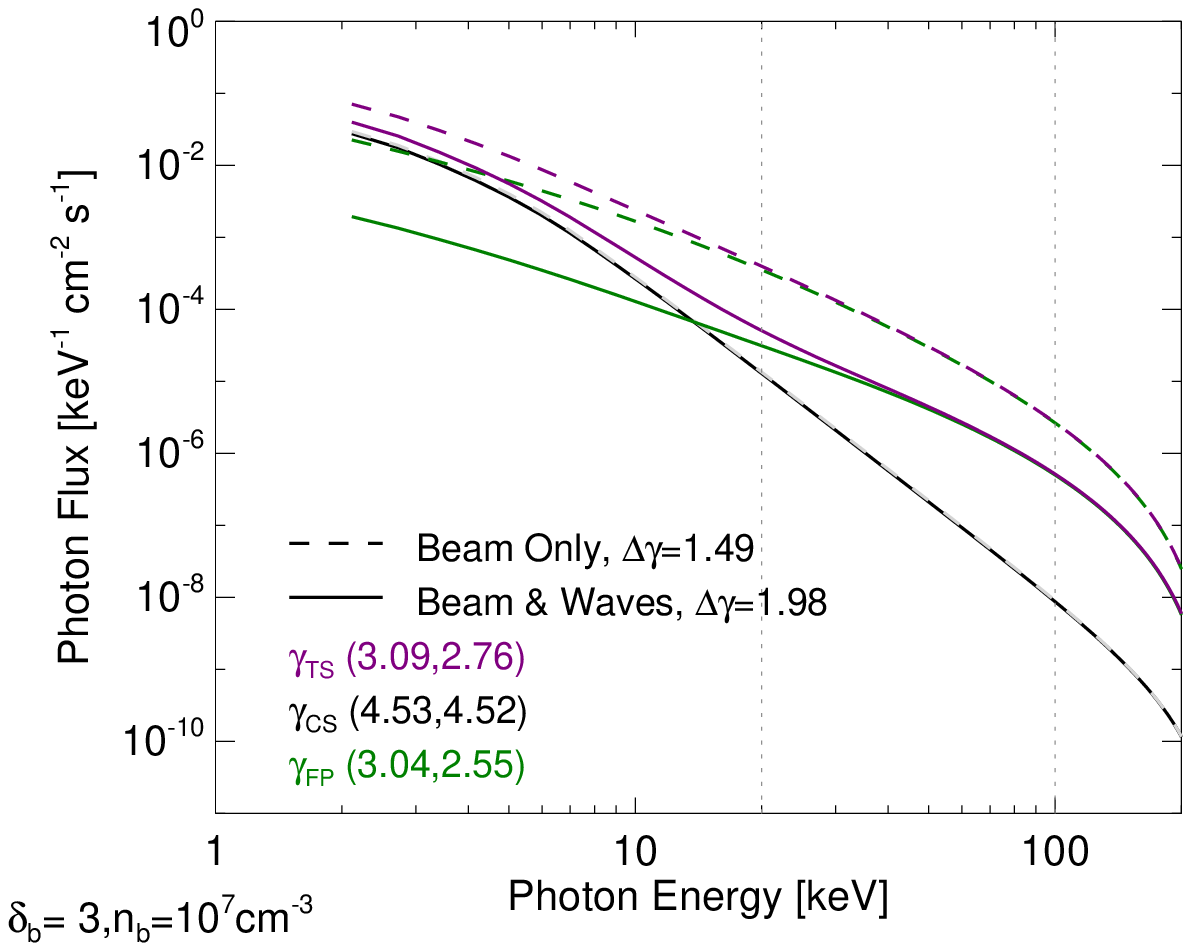}
\includegraphics[width=0.9\columnwidth]{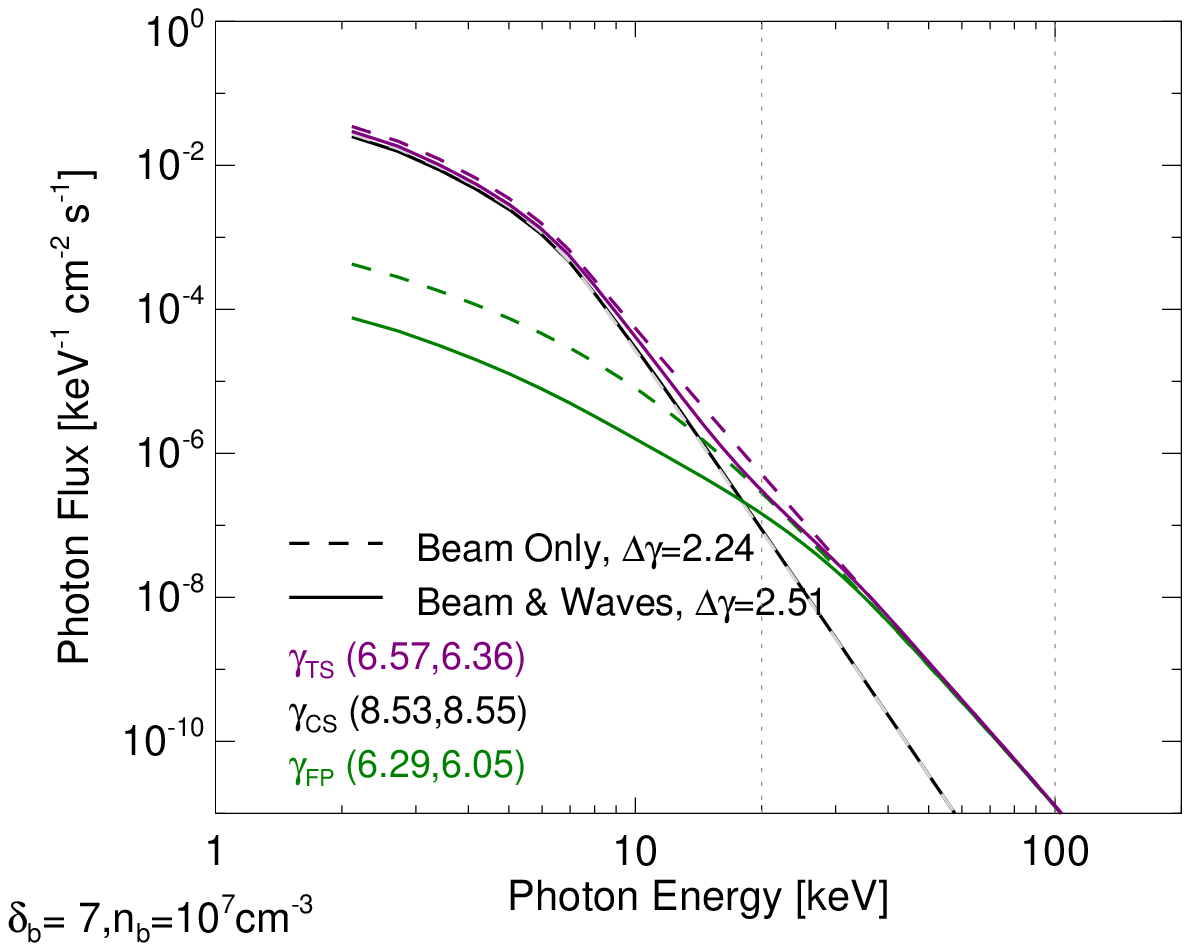}
\caption{\label{fig:spec_delta}The X-ray spectrum of an transiently injected
electron beam simulation over 1~second, for the footpoint (green), coronal
(black) and total (purple) source, with initial spectral index of
$\delta_\mathrm{b}=3$ (left) and $\delta_\mathrm{b}=7$ (right). This is in
comparison to the case of $\delta_\mathrm{b}=5$ shown in Figure
\ref{fig:inspec}.}
\end{figure*}

\begin{figure}\centering
\includegraphics[width=\columnwidth]{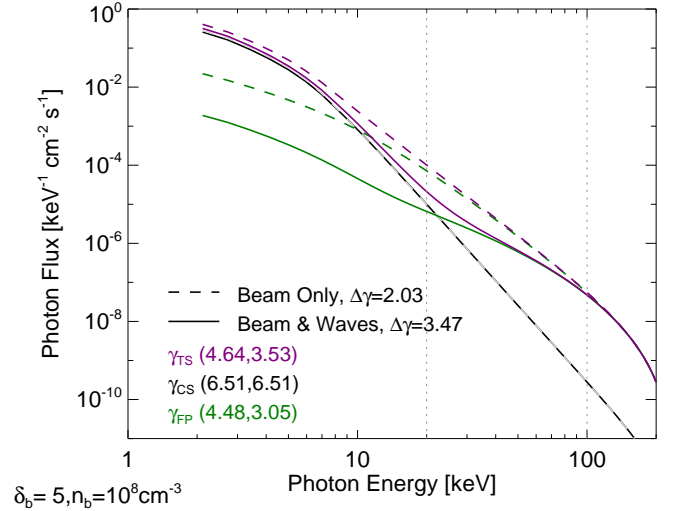}
\caption{\label{fig:spec_n}The X-ray
spectrum of an transiently injected electron beam
simulation over 1~second, for the footpoint (green), coronal
(black) and total (purple) source, with initial spectral index of
$\delta_\mathrm{b}=5$ (left) and
beam density $n_\mathrm{b}=10^{8}$ cm$^{-3}$. This is in comparison to
the case of $n_\mathrm{b}=10^{7}$ cm$^{-3}$ shown in Figure \ref{fig:inspec}.}
\end{figure}

The difference between the two cases results from considerably more wave
growth at higher velocities in the transient evolving case. This difference is
shown in Figure \ref{fig:fxv1010} where we have plotted the profile of the
electron and the wave spectral energy distribution at the height of 10~Mm in
the simulation for both the steady-state case, and for various times during the
evolving case. From the quasilinear term in Equation \ref{eq:w} we see that we
get wave growth from a positive gradient in the electron distribution $\partial
f/\partial v > 1$ as well as from the spontaneous emission where there is a
concentration of electrons, $Sf$. In the evolving case, the electrons with the
highest velocities move away from the bulk of the distribution creating a
positive gradient on the leading edge resulting in strong wave growth. At lower
velocities the emission is dominated by the spontaneous emission from the bulk
of the electron distribution but this has little effect on the X-ray spectrum as
it is
the higher velocity electrons that can greatly change the X-ray emission. If
this
is allowed to reach a steady-state with a continuously supply of electrons
behind those that move away, the wave growth is highly diminished at high
velocities, since the strong wave growth from the leading edge of the electron
distribution is washed out by the following electrons. This higher rate of wave
growth in the transiently injected case for the high energy electrons provides
an
additional mechanism for them to lose energy to the background plasma and
hence flattens both the electron and X-ray spectra.

\subsubsection{Initial Spectral Index}

In Figure \ref{fig:spec_delta} we show additional simulations of the transient
initially injected electron beam with different initial spectral indices
$\delta_\mathrm{b}=3$ (left) and $\delta_\mathrm{b}=7$ (right), instead of
$\delta_\mathrm{b}=5$ shown in Figure \ref{fig:inspec}. For the harder (flatter)
spectrum ($\delta_\mathrm{b}=3$) there is a major difference between the
beam only and the beam and waves simulations in both the spectral indices and
magnitude of emission. This is again due to there being more high energy
electrons which have an additional energy loss mechanism due to the wave
growth.

In contrast, with a steeper (harder) initial spectrum ($\delta_\mathrm{b}=7$)
there is a smaller difference between the two transport cases, as there are
fewer high energy electrons, so there is considerably lower rate of wave growth
and consequently collisional effects dominate throughout. As a result the
increase in the difference between in the spectral indices between the coronal
and footpoint sources for the beam only and beam and waves simulations is
almost half that with the flatter spectrum than with the steeper spectrum.

\begin{figure*}\centering
\includegraphics[width=0.9\columnwidth]{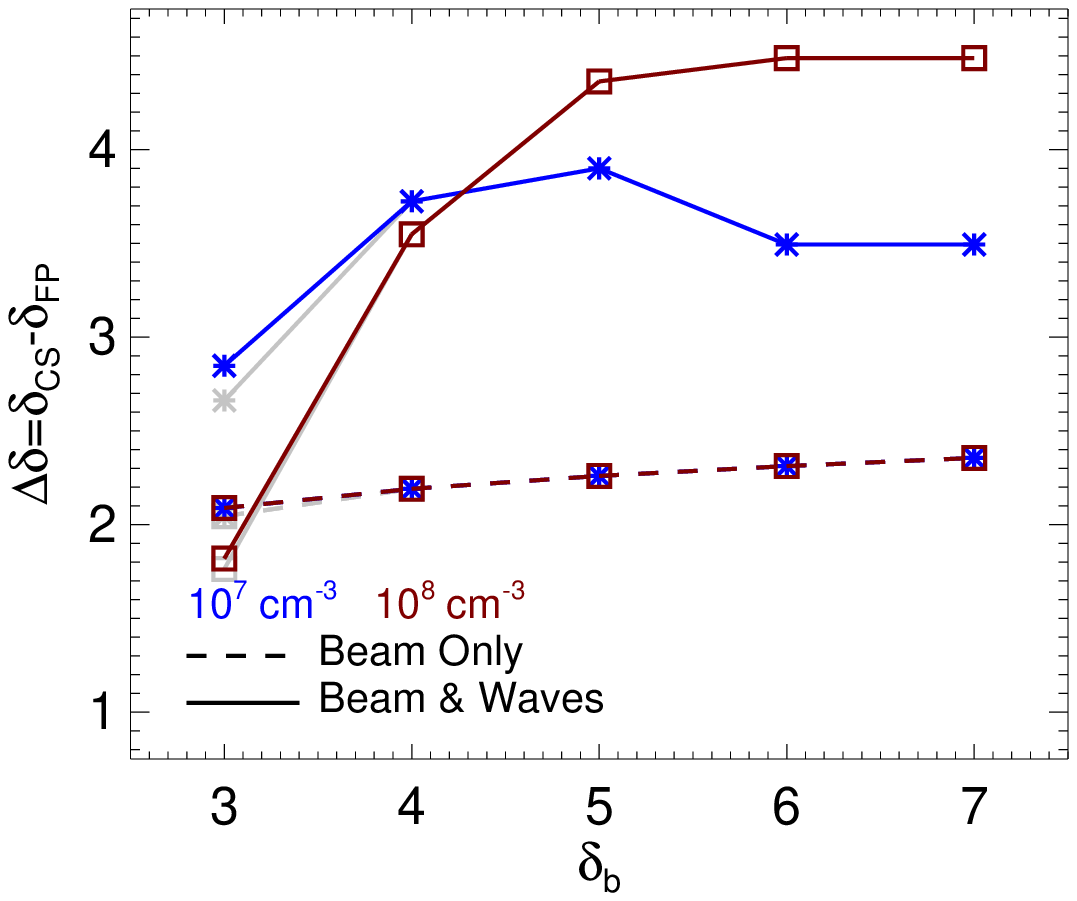}
\includegraphics[width=0.9\columnwidth]{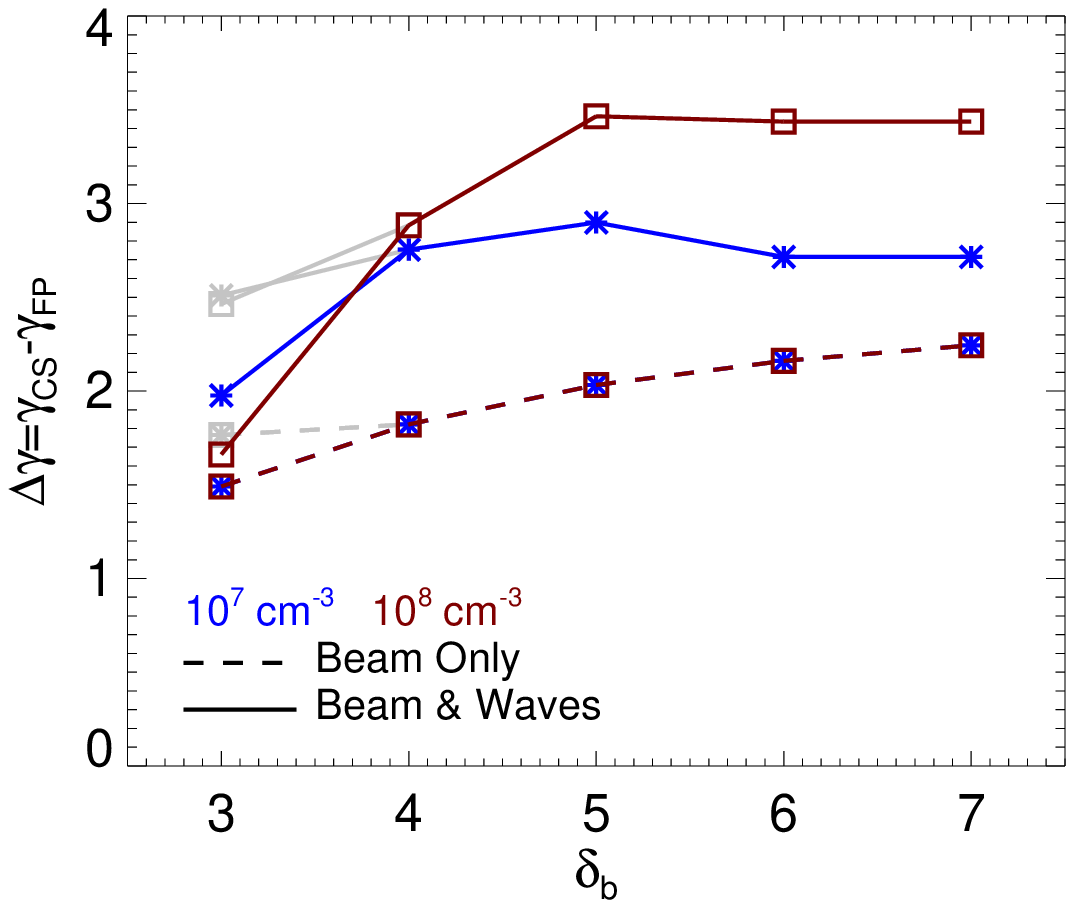}
\caption{\label{fig:a2dg}Difference in the spectral index of the coronal source
and footpoint for the mean electron spectrum $\Delta \delta$ (left) and X-ray
spectrum $\Delta \gamma$ (right) as a function on the spectral index of the
injected electron beam $\delta_\mathrm{b}$ and beam density $n_\mathrm{b}$
(blue $10^7$ cm$^{-3}$, red $10^8$ cm$^{-3}$). Shown are the results for the
beam only propagation (subject only to Coulomb collisions) and the beams and
waves simulation. The grey points indicate the results when a higher maximum
velocity is used, discussed in \S\ref{sec:maxv}.}
\end{figure*}

\subsubsection{Initial Electron Beam Density}

In Figure \ref{fig:spec_n} we have increased the initial beam density by an
order
of magnitude compared to Figure \ref{fig:inspec}, to
$n_\mathrm{b}=10^8$cm$^{-3}$. For the beam only case, the increase in the
number of accelerated electrons has only increased the magnitude of the X-ray
emission, with the spectral shape, and hence difference in spectral indices
between the coronal and foopoint sources, identical to the lower beam density
simulation  (right Figure \ref{fig:inspec}).

With the inclusion of wave-particle interactions in the beam and waves case we
not only see a change in the magnitude of emission but also in the spectral
shape, with the footpoint source flatter than with a lower beam density. This is
again due to there being more electrons a higher energies, producing higher
levels of wave growth which quickly flatten the emission. This additional energy
loss mechanism also explains the difference in magnitude of the emission
between the beam only and beam and waves.  It is important to note that the
wave-particle interaction is a highly non-linear process and so can dramatically
change the HXR spectra with only a slight change of the electron beam
parameters.

\begin{figure*}\centering
\includegraphics[width=1.8\columnwidth]{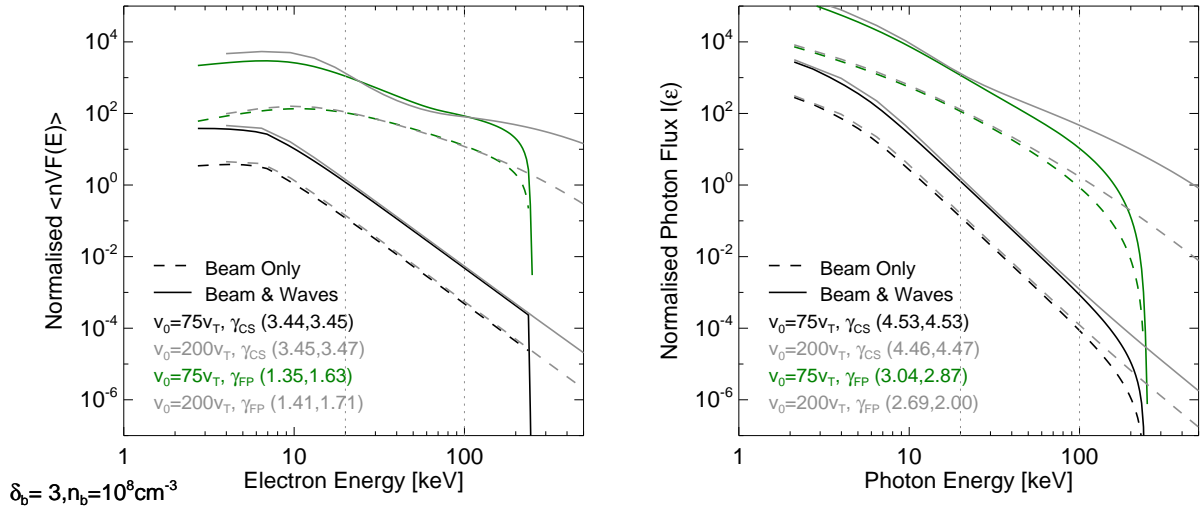}
\caption{\label{fig:spec_max}Mean electron flux spectrum (left) and X-ray
spectrum (right) of the transiently injected electron beam simulation over
1~second, for the footpoint, coronal and total source. Shown are the simulations
using a maximum velocity of $v_\mathrm{0}=75v_\mathrm{T}$
($E_\mathrm{0}=243$keV) (black coronal source, green footpoint) and
$v_\mathrm{0}=200v_\mathrm{T}$ ($E_\mathrm{0}=1730$keV) (grey lines).}
\end{figure*}

\subsection{Difference in Coronal \& Footpoint Spectral
Indices}\label{sec:diff_cf}

In Figure \ref{fig:a2dg} we show the difference in the spectral index of the
coronal and  footpoint source in the mean electron flux ($\Delta \delta$) and
X-ray ($\Delta \gamma$) spectra as a function of the initial electron
distribution
spectral index $\delta_\mathrm{b}$. Considered are both the beam only
(collisional transport, dashed lines) and beam and waves (collisions and
wave-particle interactions, solid lines) for an initial beam density of
$n_\mathrm{b}=10^7$cm$^{-3}$ and $n_\mathrm{b}=10^8$cm$^{-3}$. For the
purely collisional case we see that there is no difference as the beam density
$n_\mathrm{b}$ increases and only a small change with increasing the initial
spectral index $\delta_\mathrm{b}$. This change is due to the upper cutoff in
electron velocity (discussed further in \S \ref{sec:maxv}) and the energy range
we choose to fit over. Again we are choosing this fixed energy range as this is
comparable to the range over which a single power-law is typically fitted in HXR
imaging spectroscopy. Note that we would not expect the spectral difference to
be exactly $\Delta \gamma=2$ as this is predicted from the thin and thick-target
approximations which is closer to our steady-state simulations the transiently
injected ones presented here.

There is considerable change however with the addition of wave-particle
interactions in the beam and waves cases. The difference in the HXR spectral
indices is consistently larger than when wave-particle interactions are included
to the purely collisional case. For an increased beam density there is a bigger
difference in the HXR spectral indices  for $\delta_\mathrm{b} > 4$, typically
an
increase of around 0.7. As the initial spectral index decreases, the difference
between the different beam densities disappears (at $\delta_\mathrm{b}=4$)
and then swaps over for the flattest spectrum. The reason for this is that these
are the simulations with the largest proportion of high energy electrons in them
and hence the wave growth is having a major effect. However at these energies
we starting to approach the maximum velocity of the simulation grid, so the
results are dominated by edge effects, which we discuss in \S \ref{sec:maxv}.

\begin{figure*}\centering
\includegraphics[width=0.9\columnwidth]{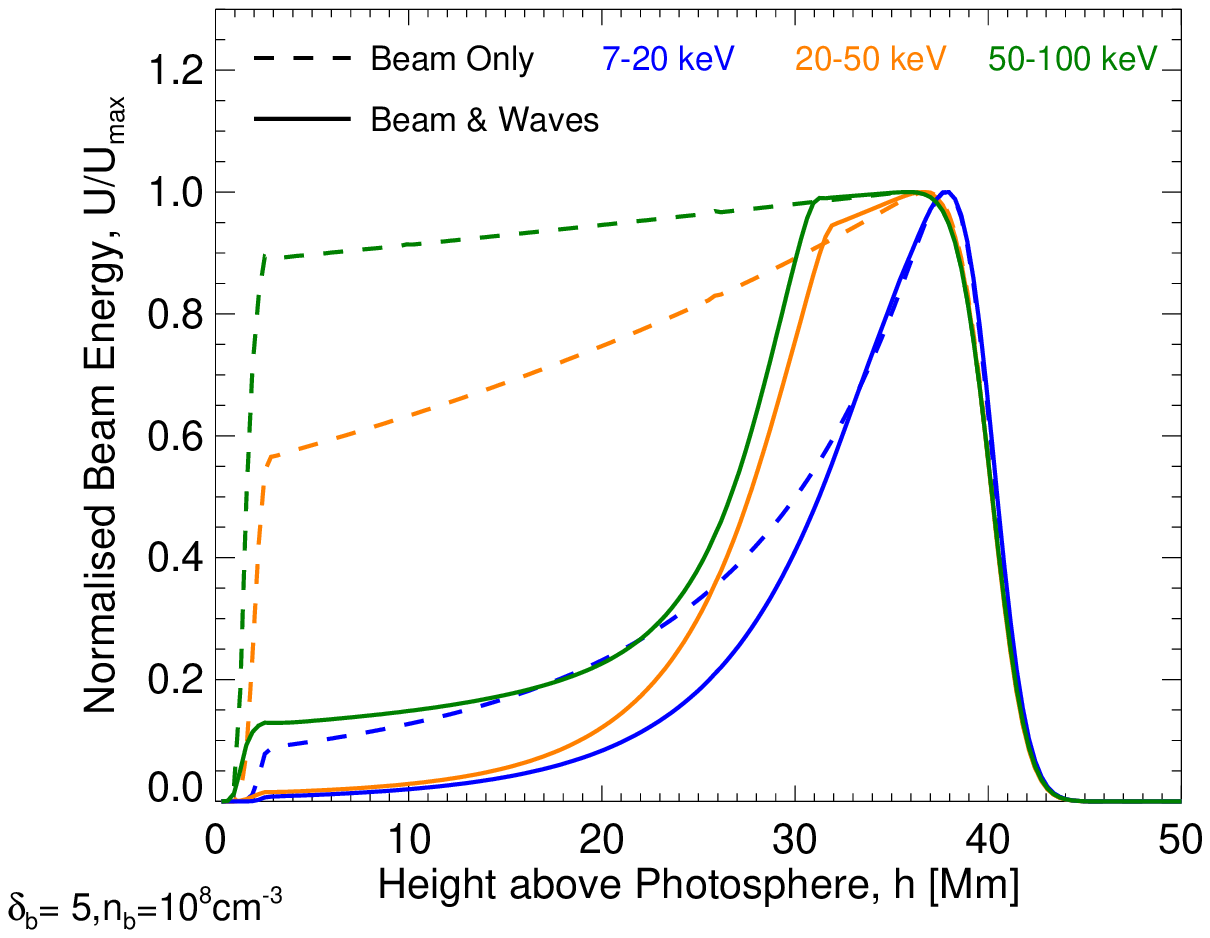}
\includegraphics[width=0.9\columnwidth]{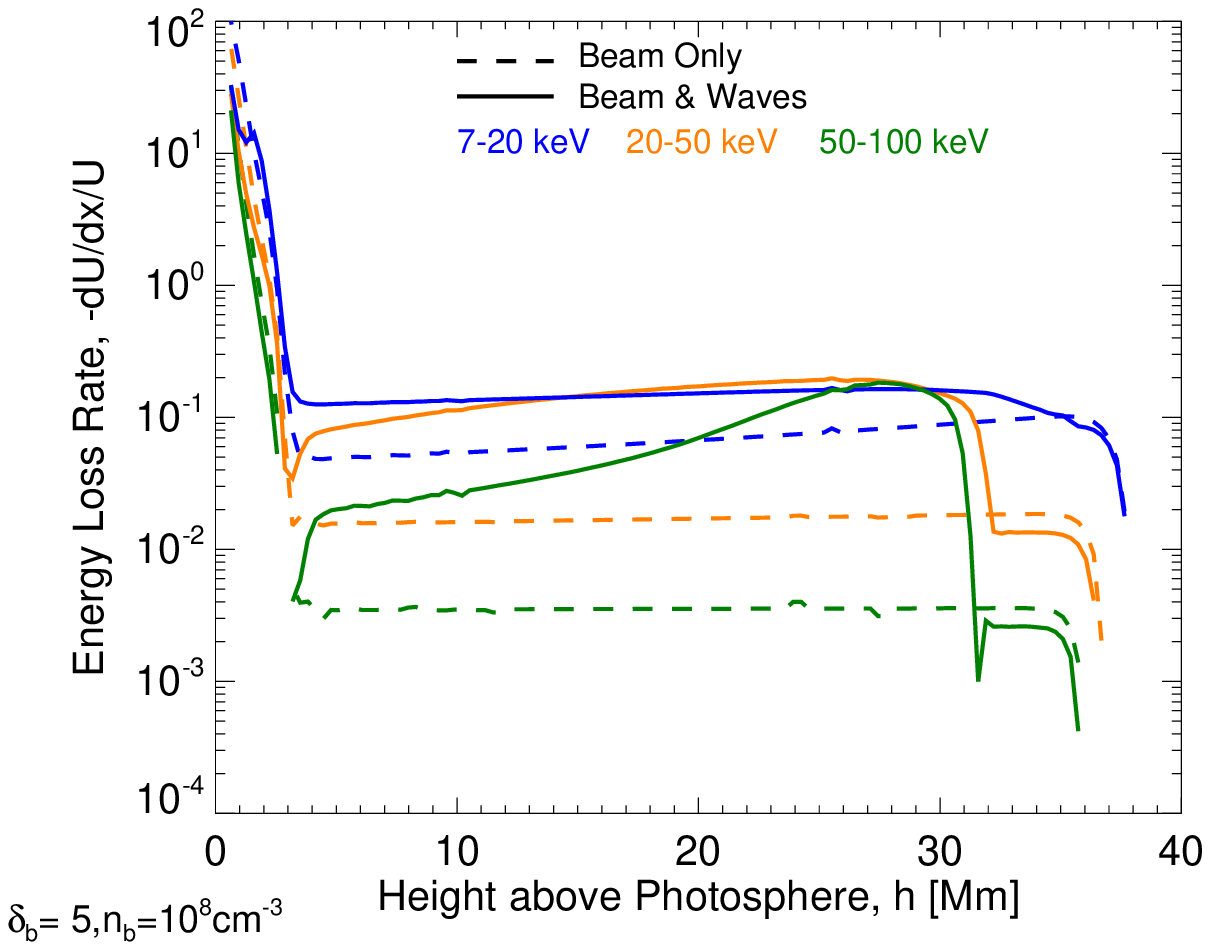}
\caption{\label{fig:loss}.(Right) The time averaged beam energy $U(x)$ for
energy bands $7-20$ keV, $20-50$ keV and $50-100$ keV (blue, orange and
green respectively) for simulation with collisional transport (dashed lines,
\emph{Beam Only}) and additional wave-particle interactions (solid lines,
\emph{Beam \& Waves}). (Left) The beam energy loss rate, normalised by the
beam energy, for the same energy bands and simulations shown in the left
panel.}
\end{figure*}

\subsubsection{Higher Maximum Velocity}\label{sec:maxv}

To investigate the role of the upper cutoff to the initial electron distribution
we
reran the most severely affected case with the most electrons initially at high
energies ($\delta_\mathrm{b}=3$ and $n_\mathrm{b}=10^8$cm$^{-3}$) with a
higher maximum velocity of $v_\mathrm{0}=200v_\mathrm{T}$
($E_\mathrm{0}=1730$keV), compared to  $v_\mathrm{0}=75v_\mathrm{T}$
($E_\mathrm{0}=243$keV) used previously. The resulting mean electron flux and
X-ray spectra for these simulations in comparison to the previous ones are
shown in Figure \ref{fig:spec_max}. In the mean electron spectra there are no
changes in the 20 to 100 keV energy range we have been fitting with a power
law for the coronal source or in the collisional case. In the footpoint source
spectrum when wave-particle interactions are included there is a slight dip
about 50 keV in the spectrum which extends to higher energies. The major
changes appear in the X-ray spectra as these are integrated over all electron
energies, as can be seen in Equation \ref{eq:brem}. The footpoint source in
particular is badly affected, with the higher cutoff producing a spectral index
that is 0.87 flatter.

Returning to Figure \ref{fig:a2dg} we can see the effect of this higher maximum
velocity in the difference in the coronal and footpoint spectral indices (grey
points). In the mean electron flux distribution there is only a minor change but
there is a substantial change in the X-ray spectral indices. With the
collisional
case the difference in spectral indices returns closer to $\Delta \gamma
\approx2$. With the wave-particle interactions we get a bigger change in the
X-ray spectral indices $2 \lesssim \Delta \gamma \lesssim 4$, similar to X-ray
observations.

Given the issue of the maximum velocity cutoff the natural question then arises
over the non-relativistic nature of our simulations. We do not feel that we need
to consider a relativistic version as we are trying to make a direct comparison
to
the collision thick-target model \citep{brown1971} which itself is
non-relativistic. Also the effect of relativism on electrons around 100 keV is
minimum (more importantly it is much less than the effect of upper
energy-cutoff) and wave-particle interactions are weaker above 100 keV.

\begin{figure*}\centering
\includegraphics[width=0.9\columnwidth]{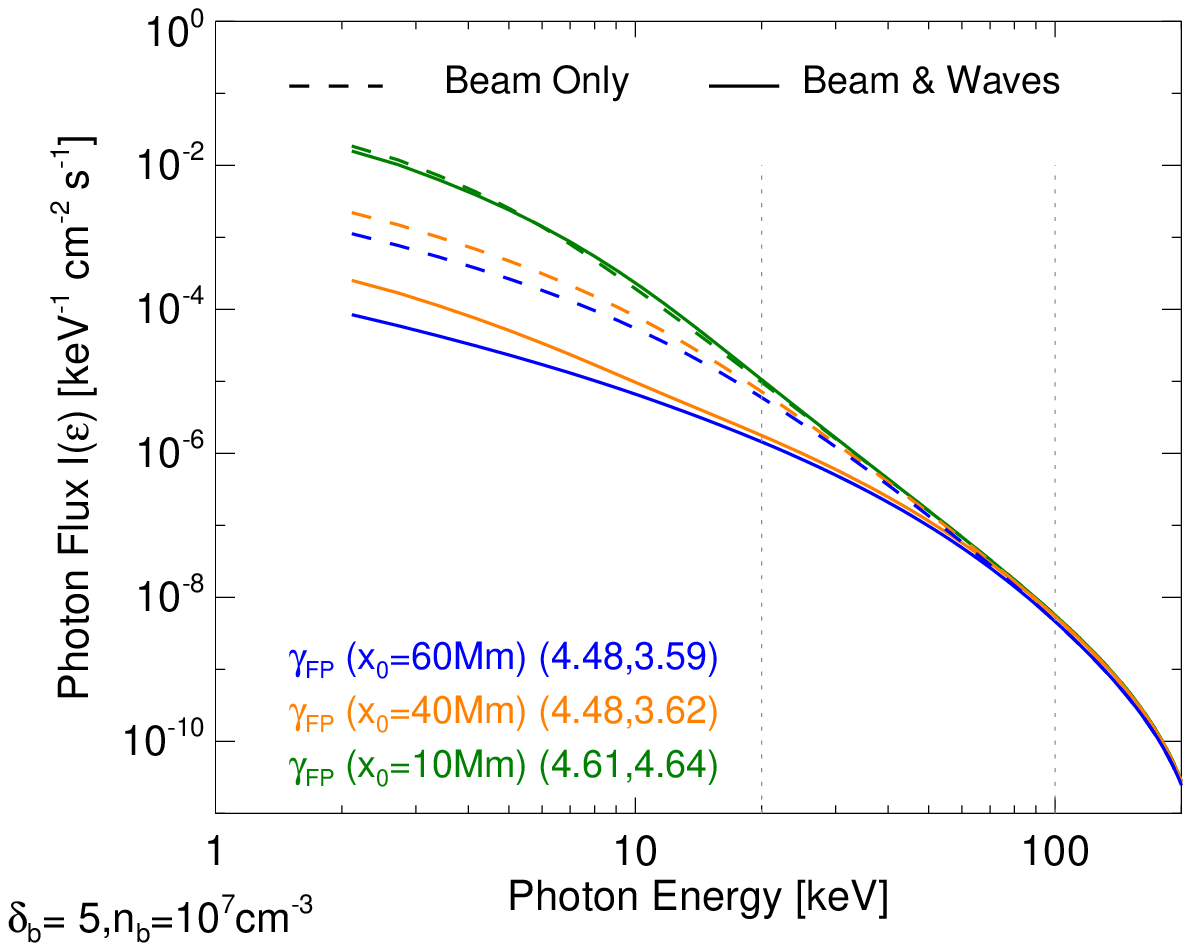}
\includegraphics[width=0.9\columnwidth]{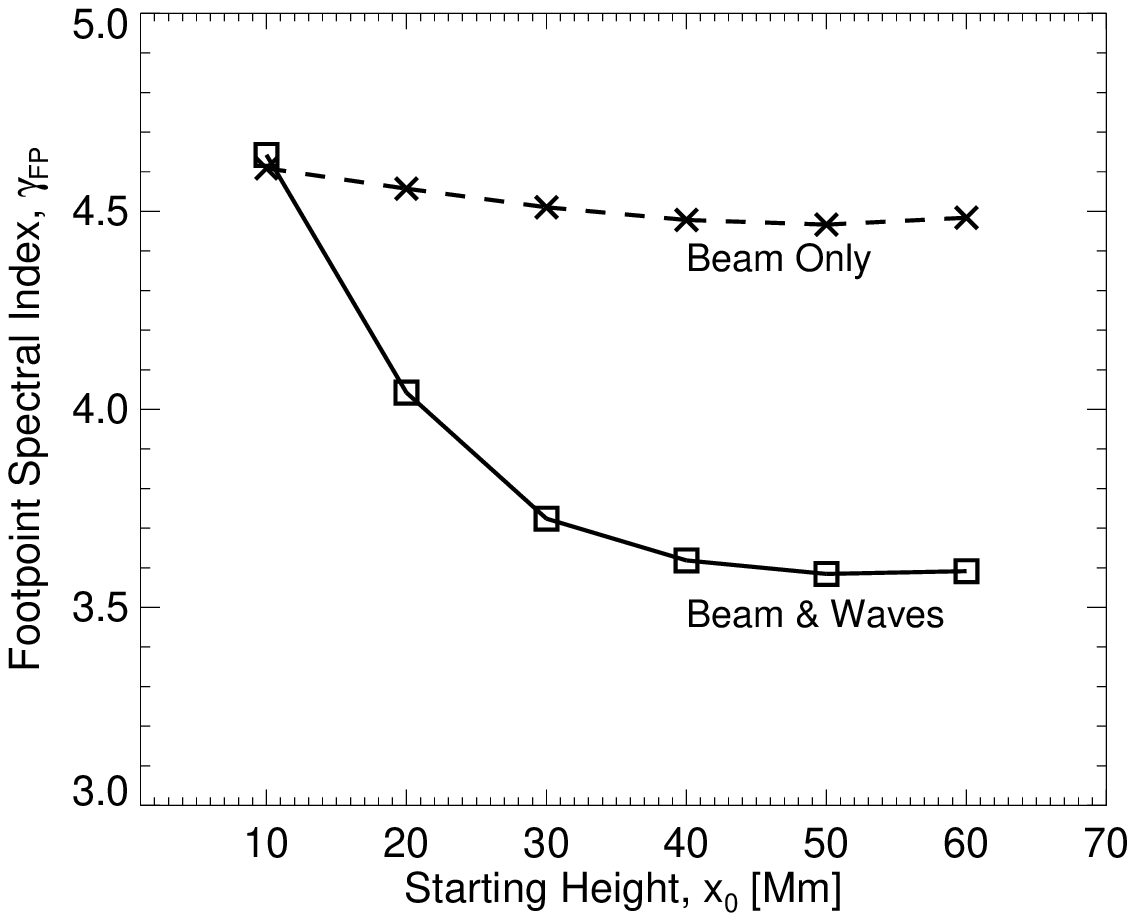}
\caption{\label{fig:x0}. (Left) HXR spectra of the footpoint region in the
simulation with $\delta_\mathrm{0}=5$, $n_\mathrm{b}=10^7$cm$^{-3}$ for
various starting heights $x_\mathrm{0}=10,40,60$ Mm (green, orange and blue
respectively) for both purely collisional and beam and wave interactions.
(Right)
The fitted spectral index over 20 to 100 keV of the footpoint region as a
function
of electron beam starting height $x_\mathrm{0}$ for purely collisional transport
(beam only) and additional wave-particle interactions.}
\end{figure*}

\subsection{Beam Energy}

The energy content of the beam is calculated as
\begin{equation}
U(x,t)=\frac{1}{2}\int_{v_1}^{v_2} v^2 \cdot f(v,x,t) dv,
\end{equation}

\noindent so variations in the beam energy at a particular time or position is
due
either to there being a change of electron velocity or number of electrons in
that
velocity/energy range. This can either occur from the particles gaining or
losing
energy to the background plasma or the beam moving through the particular
region. In Figure \ref{fig:loss} we have plotted this beam energy, time averaged
over the whole 1 second of simulation time and normalized by the maximum
energy, for various energy bands and both processes. Here we see that the
simulations with wave-particle interactions quickly lose energy in the corona.
This again matches what we have found with the flattening of the mean
electron and X-ray spectra, for example Figure \ref{fig:inspec}, where the waves
provide the electron beam an additional energy loss mechanism to the
background plasma {, heating the leg of the loop instead of the footpoints.
In the purely collisional case, the energy content of the beam between 20-50
keV and 50-100 keV is 55\% and 90\% of its initial coronal value when it reaches
the chromosphere, only 45\% and 10\% lost to heating the leg. With the
additional wave-particle interactions the beam in these energy ranges losses
the majority of its energy in the corona, with 90\% and 80\% of its initial
energy
being lost to heat the loop at heights above 20 Mm.}

We can use this time averaged beam energy to calculate the energy loss rate as
a function of spatial position $\partial U/\partial x$, shown in the righthand
side
of Figure \ref{fig:loss}. For the Coulomb collision only simulations (\emph{Beam
Only}) we see that the energy loss rate closely matches the profile of the
background plasma density: it is constant in the corona and sharply rises
through the transition region and chromosphere, i.e Figure \ref{fig:n0}. For the
cases where we have included wave-particle interactions we see a similar
structure to that of purely collisional transport only high in the corona and in
the
chromosphere. In these regions the waves have not been able to grow and the
electron transport is dominated by collisional losses respectively. In the rest
of
the corona the \emph{Beam \& Waves} case deviates from the purely collisional
simulations, the energy loss rate being far higher for the majority of the
beam's
transport in the corona.

\subsection{Starting Height: Difference in Footpoint Spectral
Indices}\label{sec:height}

Another parameter of the initial electron beam to vary is the starting height
$x_\mathrm{0}$. This is of interest as it might be able to explain the observed
difference in spectral index between double footpoint sources. In a symmetrical
injection there would be the same starting height (or path length) from
acceleration/injection site down to the footpoint. In an asymmetrical loop one
footpoint would be closer to the site of injection and the other further away.
This can be approximated in our simulations by using different starting heights
given that the coronal background density profile is the same. The results for
various starting heights are shown in Figure \ref{fig:x0}. For all heights the
result for purely collisional transport is very similar but varies when
wave-particle interactions are considered. For the lowest starting height there
is
little time for wave growth to develop to any significant level and so the
resulting electron and HXR spectrum is similar to the purely collisional case.
So
for the initial beam parameters shown we can achieve a difference in footpoint
spectral index of around $\Delta \gamma_\mathrm{FP} \approx 1$ from just
changing the starting height of the injection. This would have the largest
effect
for a loop with one leg path length from the acceleration site to footpoint of
$<
20$Mm and the other $>20$Mm: either a short $30$Mm with a small amount of
asymmetry or a highly asymmetric longer loop. Typically for microflares the SXR
thermal loop is typically $30$Mm long \citep{hannah2008} though this would be
shorter than the combined distance from the acceleration site to both
footpoints. The time of flight distance (assumed acceleration site to HXR
footpoint) was found to range between 10 and 50 Mm
\citep{1996ApJ...470.1198A}.

\section{Discussion \& Conclusions}

The simulations presented here show the importance of including wave-particle
interactions and to simulate the time and spatial evolution of flare emission.
The growth of Langmuir waves flattens the flare spectra producing a larger
spectral difference between the coronal and footpoint sources compared to
purely collisional transport. This might be able to explain the $\Delta \gamma >
2$ observed in some flares with RHESSI. Also, when considering asymmetrical
loop injection, wave-particle interactions maybe able to explain the differing
spectral indices between pairs of simultaneous footpoints without modifying the
background plasma density profile. Langmuir waves is the dominate process in
the corona and provides and additional mechanism for the beam to quickly lose
energy to the background plasma. This however does result in fewer high energy
electrons and fainter HXR spectra than those from purely collisional transport.
Note that they quickly lose their energy to a few keV and then take longer to
reach closer to the background thermal level.  {The wave-particle
interactions therefore help produce spectral indices closer to  observations but
require a larger number of injected electrons in comparison with collisional
transport.} Ways to alleviate this problem could include additional
chromospheric energisation \citep{2009A&A...508..993B} or wave refraction
from inhomogeneities in the background plasma
\citep{2001SoPh..202..131K,2010ApJ...721..864R}. Both are currently under
investigation and subject to future publications.

This work is a step towards a more complete treatment of electron transport in
solar flares with many other processes still to be considered. For instance we do
not consider the anisotropy of the electron beam or the convergence of the
magnetic field and simulate only one spatial dimension. The Langmuir waves
generated in our simulations are the seeds for radio emission which are
sometimes seen in reverse type III bursts. Therefore, a natural next step for this
work would be to calculate the radio emission from the propagating electron
beam. This is a complicated task given the highly non-linear processes involved
in the emission. By using both radio and highly sensitive HXR observations in
comparison to simulations such as those presented in this paper we would be
able to greatly constrain the nature of particle acceleration and transport in
solar flares.

The HXR imaging spectroscopy of RHESSI has shown the importance of such
observations for solar flare physics. However to really pin down the acceleration
and transport processes would require having solar HXR imaging spectroscopy
instruments providing orders of magnitude better sensitivity and dynamic range
over RHESSI, so that the coronal source emission is routinely observed with the
bright footpoint source. Such advances in sensitivity and dynamic range are
achievable with HXR focusing optics such as those on FOXSI
\citep{2009SPIE.7437E...4K} and NuSTAR \citep{2010SPIE.7732E..21H}.

\begin{acknowledgements}

We thank the referee for their constructive comments. This work is supported by
a STFC rolling grant (IGH, EPK) and STFC Advanced Fellowship (EPK). Financial
support by the Royal Society (Research Grant) and the European Commission
through the SOLAIRE Network (MTRN-CT-2006-035484) is gratefully
acknowledged.

\end{acknowledgements}


\begin{thebibliography}{59}
\expandafter\ifx\csname natexlab\endcsname\relax\def\natexlab#1{#1}\fi

\bibitem[{{Alexander} \& {Metcalf}(1997)}]{1997ApJ...489..442A}
{Alexander}, D. \& {Metcalf}, T.~R. 1997, \apj, 489, 442

\bibitem[{{Arnoldy} {et~al.}(1968){Arnoldy}, {Kane}, \&
  {Winckler}}]{1968ApJ...151..711A}
{Arnoldy}, R.~L., {Kane}, S.~R., \& {Winckler}, J.~R. 1968, \apj, 151, 711

\bibitem[{{Aschwanden} \& {Benz}(1997)}]{1997ApJ...480..825A}
{Aschwanden}, M.~J. \& {Benz}, A.~O. 1997, \apj, 480, 825

\bibitem[{{Aschwanden} {et~al.}(1995){Aschwanden}, {Benz}, {Dennis}, \&
  {Schwartz}}]{1995ApJ...455..347A}
{Aschwanden}, M.~J., {Benz}, A.~O., {Dennis}, B.~R., \& {Schwartz}, R.~A. 1995,
  \apj, 455, 347

\bibitem[{{Aschwanden} {et~al.}(1998){Aschwanden}, {Kliem}, {Schwarz},
  {Kurths}, {Dennis}, \& {Schwartz}}]{Aschwanden_etal1998}
{Aschwanden}, M.~J., {Kliem}, B., {Schwarz}, U., {et~al.} 1998, \apj, 505, 941

\bibitem[{{Aschwanden} {et~al.}(1996){Aschwanden}, {Kosugi}, {Hudson}, {Wills},
  \& {Schwartz}}]{1996ApJ...470.1198A}
{Aschwanden}, M.~J., {Kosugi}, T., {Hudson}, H.~S., {Wills}, M.~J., \&
  {Schwartz}, R.~A. 1996, \apj, 470, 1198

\bibitem[{{Battaglia} \& {Benz}(2006)}]{2006A&A...456..751B}
{Battaglia}, M. \& {Benz}, A.~O. 2006, \aap, 456, 751

\bibitem[{{Battaglia} \& {Benz}(2007)}]{2007A&A...466..713B}
{Battaglia}, M. \& {Benz}, A.~O. 2007, \aap, 466, 713

\bibitem[{{Battaglia} \& {Benz}(2008)}]{2008A&A...487..337B}
{Battaglia}, M. \& {Benz}, A.~O. 2008, \aap, 487, 337

\bibitem[{{Benz} {et~al.}(2005){Benz}, {Grigis}, {Csillaghy}, \&
  {Saint-Hilaire}}]{2005SoPh..226..121B}
{Benz}, A.~O., {Grigis}, P.~C., {Csillaghy}, A., \& {Saint-Hilaire}, P. 2005,
  \solphys, 226, 121

\bibitem[{{Brown}(1971)}]{brown1971}
{Brown}, J.~C. 1971, \solphys, 18, 489

\bibitem[{{Brown} {et~al.}(2006){Brown}, {Emslie}, {Holman}, {Johns-Krull},
  {Kontar}, {Lin}, {Massone}, \& {Piana}}]{Brown_etal2006}
{Brown}, J.~C., {Emslie}, A.~G., {Holman}, G.~D., {et~al.} 2006, \apj, 643, 523

\bibitem[{{Brown} {et~al.}(2003){Brown}, {Emslie}, \&
  {Kontar}}]{2003ApJ...595L.115B}
{Brown}, J.~C., {Emslie}, A.~G., \& {Kontar}, E.~P. 2003, \apjl, 595, L115

\bibitem[{{Brown} {et~al.}(2009){Brown}, {Turkmani}, {Kontar}, {MacKinnon}, \&
  {Vlahos}}]{2009A&A...508..993B}
{Brown}, J.~C., {Turkmani}, R., {Kontar}, E.~P., {MacKinnon}, A.~L., \&
  {Vlahos}, L. 2009, \aap, 508, 993

\bibitem[{{Drummond} \& {Pines}(1964)}]{DrummondPines1964}
{Drummond}, W.~E. \& {Pines}, D. 1964, Annals of Physics, 28, 478

\bibitem[{{Emslie}(1978)}]{1978ApJ...224..241E}
{Emslie}, A.~G. 1978, \apj, 224, 241

\bibitem[{{Emslie} {et~al.}(2003){Emslie}, {Kontar}, {Krucker}, \&
  {Lin}}]{2003ApJ...595L.107E}
{Emslie}, A.~G., {Kontar}, E.~P., {Krucker}, S., \& {Lin}, R.~P. 2003, \apjl,
  595, L107

\bibitem[{{Emslie} \& {Smith}(1984)}]{1984ApJ...279..882E}
{Emslie}, A.~G. \& {Smith}, D.~F. 1984, \apj, 279, 882

\bibitem[{{Feldman} {et~al.}(1994){Feldman}, {Hiei}, {Phillips}, {Brown}, \&
  {Lang}}]{1994ApJ...421..843F}
{Feldman}, U., {Hiei}, E., {Phillips}, K.~J.~H., {Brown}, C.~M., \& {Lang}, J.
  1994, \apj, 421, 843

\bibitem[{{Fleishman} {et~al.}(1994){Fleishman}, {Stepanov}, \&
  {Yurovsky}}]{1994SoPh..153..403F}
{Fleishman}, G.~D., {Stepanov}, A.~V., \& {Yurovsky}, Y.~F. 1994, \solphys,
  153, 403

\bibitem[{{Hamilton} \& {Petrosian}(1987)}]{hamilton1987}
{Hamilton}, R.~J. \& {Petrosian}, V. 1987, \apj, 321, 721

\bibitem[{{Hannah} {et~al.}(2008){Hannah}, {Christe}, {Krucker}, {Hurford},
  {Hudson}, \& {Lin}}]{hannah2008}
{Hannah}, I.~G., {Christe}, S., {Krucker}, S., {et~al.} 2008, \apj, 677, 704

\bibitem[{{Hannah} {et~al.}(2009){Hannah}, {Kontar}, \&
  {Sirenko}}]{2009ApJ...707L..45H}
{Hannah}, I.~G., {Kontar}, E.~P., \& {Sirenko}, O.~K. 2009, \apjl, 707, L45

\bibitem[{{Harrison} {et~al.}(2010){Harrison}, {Boggs}, {Christensen}, {Craig},
  {Hailey}, {Stern}, {Zhang}, {Angelini}, {An}, {Bhalereo}, {Brejnholt},
  {Cominsky}, {Cook}, {Doll}, {Giommi}, {Grefenstette}, {Hornstrup}, {Kaspi},
  {Kim}, {Kitaguchi}, {Koglin}, {Liebe}, {Madejski}, {Kruse Madsen}, {Mao},
  {Meier}, {Miyasaka}, {Mori}, {Perri}, {Pivovaroff}, {Puccetti}, {Rana}, \&
  {Zoglauer}}]{2010SPIE.7732E..21H}
{Harrison}, F.~A., {Boggs}, S., {Christensen}, F., {et~al.} 2010, in Society of
  Photo-Optical Instrumentation Engineers (SPIE) Conference Series, Vol. 7732,
  Society of Photo-Optical Instrumentation Engineers (SPIE) Conference Series

\bibitem[{{Haug}(1997)}]{1997A&A...326..417H}
{Haug}, E. 1997, \aap, 326, 417

\bibitem[{{Holt} \& {Cline}(1968)}]{1968ApJ...154.1027H}
{Holt}, S.~S. \& {Cline}, T.~L. 1968, \apj, 154, 1027

\bibitem[{{Karlick{\'y}} \& {Ka{\v s}parov{\'a}}(2009)}]{2009A&A...506.1437K}
{Karlick{\'y}}, M. \& {Ka{\v s}parov{\'a}}, J. 2009, \aap, 506, 1437

\bibitem[{{Kiplinger} {et~al.}(1984){Kiplinger}, {Dennis}, {Frost}, \&
  {Orwig}}]{Kiplinger_etal1984}
{Kiplinger}, A.~L., {Dennis}, B.~R., {Frost}, K.~J., \& {Orwig}, L.~E. 1984,
  \apjl, 287, L105

\bibitem[{{Klein} {et~al.}(1997){Klein}, {Aurass}, {Soru-Escaut}, \&
  {Kalman}}]{1997A&A...320..612K}
{Klein}, K.-L., {Aurass}, H., {Soru-Escaut}, I., \& {Kalman}, B. 1997, \aap,
  320, 612

\bibitem[{{Koch} \& {Motz}(1959)}]{1959RvMP...31..920K}
{Koch}, H.~W. \& {Motz}, J.~W. 1959, Reviews of Modern Physics, 31, 920

\bibitem[{{Kontar}(2001{\natexlab{a}})}]{2001SoPh..202..131K}
{Kontar}, E.~P. 2001{\natexlab{a}}, \solphys, 202, 131

\bibitem[{{Kontar}(2001{\natexlab{b}})}]{2001CoPhC.138..222K}
{Kontar}, E.~P. 2001{\natexlab{b}}, Computer Physics Communications, 138, 222

\bibitem[{{Kontar} {et~al.}(2010){Kontar}, {Hannah}, {Jeffrey}, \&
  {Battaglia}}]{2010ApJ...717..250K}
{Kontar}, E.~P., {Hannah}, I.~G., {Jeffrey}, N.~L.~S., \& {Battaglia}, M. 2010,
  \apj, 717, 250

\bibitem[{{Kontar} {et~al.}(2008){Kontar}, {Hannah}, \&
  {MacKinnon}}]{2008A&A...489L..57K}
{Kontar}, E.~P., {Hannah}, I.~G., \& {MacKinnon}, A.~L. 2008, \aap, 489, L57

\bibitem[{{Kosugi} {et~al.}(1992){Kosugi}, {Sakao}, {Masuda}, {Makishima},
  {Inda}, {Murakami}, {Ogawara}, {Yaji}, \& {Matsushita}}]{1992PASJ...44L..45K}
{Kosugi}, T., {Sakao}, T., {Masuda}, S., {et~al.} 1992, \pasj, 44, L45

\bibitem[{{Krucker} {et~al.}(2009){Krucker}, {Christe}, {Glesener}, {McBride},
  {Turin}, {Glaser}, {Saint-Hilaire}, {Delory}, {Lin}, {Gubarev}, {Ramsey},
  {Terada}, {Ishikawa}, {Kokubun}, {Saito}, {Takahashi}, {Watanabe},
  {Nakazawa}, {Tajima}, {Masuda}, {Minoshima}, \&
  {Shomojo}}]{2009SPIE.7437E...4K}
{Krucker}, S., {Christe}, S., {Glesener}, L., {et~al.} 2009, in Society of
  Photo-Optical Instrumentation Engineers (SPIE) Conference Series, Vol. 7437,
  Society of Photo-Optical Instrumentation Engineers (SPIE) Conference Series

\bibitem[{{Krucker} {et~al.}(2010){Krucker}, {Hudson}, {Glesener}, {White},
  {Masuda}, {Wuelser}, \& {Lin}}]{2010ApJ...714.1108K}
{Krucker}, S., {Hudson}, H.~S., {Glesener}, L., {et~al.} 2010, \apj, 714, 1108

\bibitem[{{Lifshitz} \& {Pitaevskii}(1981)}]{1981phki.book.....L}
{Lifshitz}, E.~M. \& {Pitaevskii}, L.~P. 1981, {Physical kinetics} (Course of
  theoretical physics, Oxford: Pergamon Press, 1981)

\bibitem[{{Lin} {et~al.}(2002){Lin}, {Dennis}, {Hurford}, {Smith}, {Zehnder},
  {Harvey}, {Curtis}, {Pankow}, {Turin}, {Bester}, {Csillaghy}, {Lewis},
  {Madden}, {van Beek}, {Appleby}, {Raudorf}, {McTiernan}, {Ramaty}, {Schmahl},
  {Schwartz}, {Krucker}, {Abiad}, {Quinn}, {Berg}, {Hashii}, {Sterling},
  {Jackson}, {Pratt}, {Campbell}, {Malone}, {Landis}, {Barrington-Leigh},
  {Slassi-Sennou}, {Cork}, {Clark}, {Amato}, {Orwig}, {Boyle}, {Banks},
  {Shirey}, {Tolbert}, {Zarro}, {Snow}, {Thomsen}, {Henneck}, {McHedlishvili},
  {Ming}, {Fivian}, {Jordan}, {Wanner}, {Crubb}, {Preble}, {Matranga}, {Benz},
  {Hudson}, {Canfield}, {Holman}, {Crannell}, {Kosugi}, {Emslie}, {Vilmer},
  {Brown}, {Johns-Krull}, {Aschwanden}, {Metcalf}, \&
  {Conway}}]{2002SoPh..210....3L}
{Lin}, R.~P., {Dennis}, B.~R., {Hurford}, G.~J., {et~al.} 2002, \solphys, 210,
  3

\bibitem[{{Lin} \& {Hudson}(1976)}]{1976SoPh...50..153L}
{Lin}, R.~P. \& {Hudson}, H.~S. 1976, \solphys, 50, 153

\bibitem[{{Masuda} {et~al.}(1995){Masuda}, {Kosugi}, {Hara}, {Sakao},
  {Shibata}, \& {Tsuneta}}]{1995PASJ...47..677M}
{Masuda}, S., {Kosugi}, T., {Hara}, H., {et~al.} 1995, \pasj, 47, 677

\bibitem[{{Masuda} {et~al.}(1994){Masuda}, {Kosugi}, {Hara}, {Tsuneta}, \&
  {Ogawara}}]{1994Natur.371..495M}
{Masuda}, S., {Kosugi}, T., {Hara}, H., {Tsuneta}, S., \& {Ogawara}, Y. 1994,
  \nat, 371, 495

\bibitem[{{Masuda} {et~al.}(2000){Masuda}, {Sato}, {Kosugi}, \&
  {Sakao}}]{2000AdSpR..26..493M}
{Masuda}, S., {Sato}, J., {Kosugi}, T., \& {Sakao}, T. 2000, Advances in Space
  Research, 26, 493

\bibitem[{{McClements}(1987)}]{McClements1987}
{McClements}, K.~G. 1987, \aap, 175, 255

\bibitem[{{Melrose}(1980)}]{1980panp.book.....M}
{Melrose}, D.~B. 1980, {Plasma astrohysics. Nonthermal processes in diffuse
  magnetized plasmas.} (New York: Gordon and Breach, 1980)

\bibitem[{{Melrose} \& {Brown}(1976)}]{1976MNRAS.176...15M}
{Melrose}, D.~B. \& {Brown}, J.~C. 1976, \mnras, 176, 15

\bibitem[{{Petrosian} {et~al.}(2002){Petrosian}, {Donaghy}, \&
  {McTiernan}}]{2002ApJ...569..459P}
{Petrosian}, V., {Donaghy}, T.~Q., \& {McTiernan}, J.~M. 2002, \apj, 569, 459

\bibitem[{{Reid} \& {Kontar}(2010)}]{2010ApJ...721..864R}
{Reid}, H.~A.~S. \& {Kontar}, E.~P. 2010, \apj, 721, 864

\bibitem[{{Ryutov}(1969)}]{1969JETP...30..131R}
{Ryutov}, D.~D. 1969, Soviet Journal of Experimental and Theoretical Physics,
  30, 131

\bibitem[{{Saint-Hilaire} {et~al.}(2009){Saint-Hilaire}, {Krucker}, {Christe},
  \& {Lin}}]{2009ApJ...696..941S}
{Saint-Hilaire}, P., {Krucker}, S., {Christe}, S., \& {Lin}, R.~P. 2009, \apj,
  696, 941

\bibitem[{{Saint-Hilaire} {et~al.}(2008){Saint-Hilaire}, {Krucker}, \&
  {Lin}}]{2008SoPh..250...53S}
{Saint-Hilaire}, P., {Krucker}, S., \& {Lin}, R.~P. 2008, \solphys, 250, 53

\bibitem[{{Schmahl} {et~al.}(2006){Schmahl}, {Pernak}, \&
  {Hurford}}]{2006SPD....37.1308S}
{Schmahl}, E.~J., {Pernak}, R., \& {Hurford}, G. 2006, in Bulletin of the
  American Astronomical Society, Vol.~38, Bulletin of the American Astronomical
  Society, 241--+

\bibitem[{{Su} {et~al.}(2009){Su}, {Holman}, {Dennis}, {Tolbert}, \&
  {Schwartz}}]{2009ApJ...705.1584S}
{Su}, Y., {Holman}, G.~D., {Dennis}, B.~R., {Tolbert}, A.~K., \& {Schwartz},
  R.~A. 2009, \apj, 705, 1584

\bibitem[{{Syrovatskii} \& {Shmeleva}(1972)}]{1972SvA....16..273S}
{Syrovatskii}, S.~I. \& {Shmeleva}, O.~P. 1972, \sovast, 16, 273

\bibitem[{{Takakura}(1969)}]{1969SoPh....6..133T}
{Takakura}, T. 1969, \solphys, 6, 133

\bibitem[{{Tarnstrom} \& {Zehntner}(1975)}]{1975Natur.258..693T}
{Tarnstrom}, G.~L. \& {Zehntner}, C. 1975, \nat, 258, 693

\bibitem[{{Tsytovich} \& {Terhaar}(1995)}]{Tsytovich1995}
{Tsytovich}, V.~N. \& {Terhaar}, D. 1995, {Lectures on Non-linear Plasma
  Kinetics}, ed. V.~N. {Tsytovich} \& D.~{Terhaar}

\bibitem[{{Vedenov} \& {Velikhov}(1963)}]{VedenovVelikhov1963}
{Vedenov}, A.~A. \& {Velikhov}, E.~P. 1963, Soviet Journal of Experimental and
  Theoretical Physics, 16, 682

\bibitem[{{Wheatland} \& {Melrose}(1995)}]{1995SoPh..158..283W}
{Wheatland}, M.~S. \& {Melrose}, D.~B. 1995, \solphys, 158, 283

\end{thebibliography}

\end{document}